\documentclass[aps,prb,amsmath,amssymb,reprint,superscriptaddress,preprintnumbers,showpacs,intlimits,longbibliography]{revtex4-1}
\pdfoutput=1
\bibpunct{[}{]}{,}{n}{}{}

\usepackage[german,english]{babel}
\usepackage{bm,latexsym,mathrsfs,enumerate}
\usepackage{xcolor}
\usepackage{mathtools}
\usepackage[mathcal]{euscript}
\usepackage[breaklinks=true,unicode=true,urlcolor = blue,colorlinks = true,citecolor = blue,linkcolor = blue]{hyperref}
\usepackage{graphicx}
\usepackage{wasysym}

\usepackage[labelformat=simple]{subcaption}

\usepackage[normalem]{ulem}
\usepackage[toc,page]{appendix}

\renewcommand{\vec}[1]{\bm{#1}}
\begin{document}	
	\title{Geometry induced phase transitions in magnetic spherical shell}
	
	\author{Mykola I. Sloika}
	\email{sloika.m@gmail.com}
	\affiliation{Taras Shevchenko National University of Kyiv, 01601 Kyiv, Ukraine}
	
	\author{Denis D. Sheka}
	\email{sheka@univ.net.ua}
	\affiliation{Taras Shevchenko National University of Kyiv, 01601 Kyiv, Ukraine}
	
	\author{Volodymyr~P.~Kravchuk}
	\email{vkravchuk@bitp.kiev.ua}
	\affiliation{Bogolyubov Institute for Theoretical Physics of National Academy of Sciences of Ukraine, 03680 Kyiv, Ukraine}
	
	\author{Oleksandr V. Pylypovskyi}
	\email{engraver@knu.ua}
	\affiliation{Taras Shevchenko National University of Kyiv, 01601 Kyiv, Ukraine}
	
	\author{Yuri~Gaididei}
	\email{ybg@bitp.kiev.ua}
	\affiliation{Bogolyubov Institute for Theoretical Physics of National Academy of Sciences of Ukraine, 03680 Kyiv, Ukraine}
	
	\date{October 18, 2016}
	
	%
	%
	
	\begin{abstract}
		Equilibrium magnetization states in thin spherical shells of a magnetically soft ferromagnet are determined by the competition between two interactions: (i) The local exchange interaction favors the more homogeneous onion state with magnetization oriented in meridian directions; such a state is realized in relatively small particles. (ii) The nonlocal magnetostatic interaction prefers the double-vortex configuration with the magnetization oriented in the parallels directions, since it minimizes the volume magnetostatic charges.  These states are topologically equivalent, in contrast to the same-name states of magnetic nanoring.  As a consequence, a continuous (the second order) phase transition between the vortex and onion states takes place. The detailed analytical description of the phase diagram is well confirmed by micromagnetic simulations.
	\end{abstract}

	
	%
	%
	%
	
	
	\maketitle

	
	\section{Introduction}
	\label{sec:introduction}
	
	Topological magnetization structures provide new properties to hosting nanomaterials, attracting intensive fundamental research as well as numerous applications to processing~\cite{Allwood05,Kim10,Fert13,Krause16} and information-storage devices~\cite{Cowburn07,Bohlens08,Parkin08,Kim10,Pigeau10,Yu11a,Hertel13}. Examples include domain walls~\cite{Hubert09}, vortices~\cite{Hubert09,Guimaraes09}, skyrmions~ \cite{Seki16,Seidel16}. 
	
\begin{figure}
		\begin{subfigure}{0.99\columnwidth}
			\caption{Ring: $\pi_1$-topological charge q (vorticity)}
			\label{fig:2D-ring}
			\includegraphics [width=.95\columnwidth,angle=0]{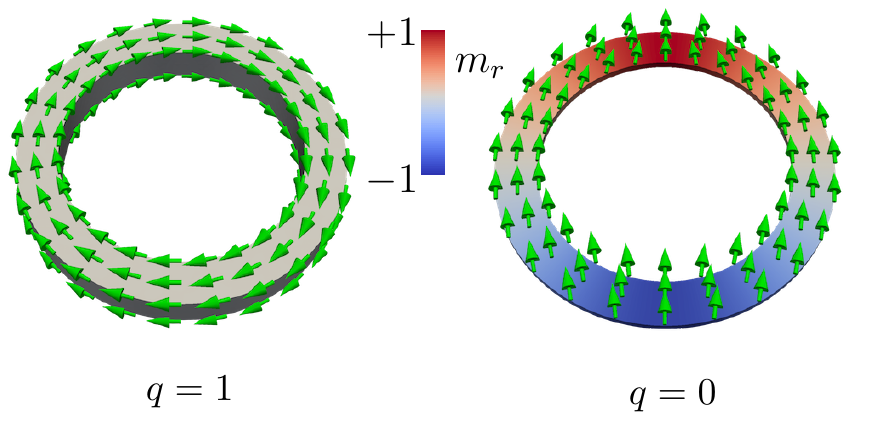}
		\end{subfigure}
		\vskip\baselineskip
		\begin{subfigure}{0.99\columnwidth}
			\caption{Sphere: $\pi_2$-topological charge Q (skyrmion number)}
			\label{fig:3D-shell}
			\includegraphics [width=.95\columnwidth,angle=0]{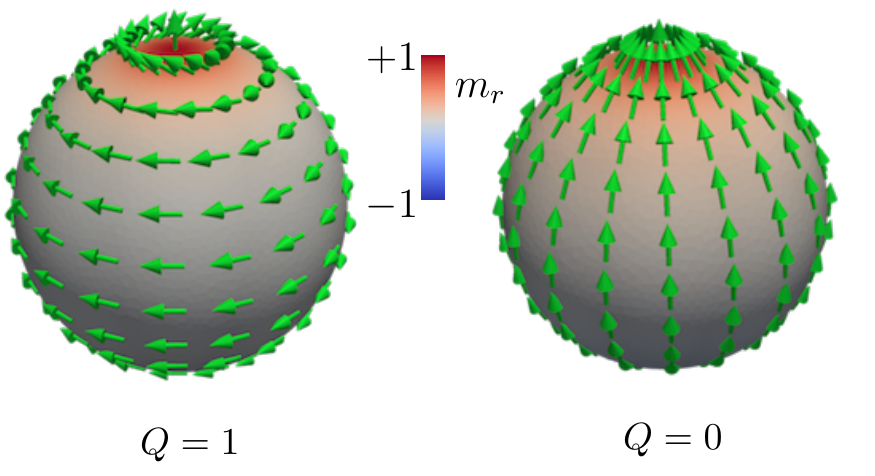}
		\end{subfigure}
		\caption{(Color online) \textbf{Topology \emph{vs} geometry:} the vortex and onion states in the case of a planar ring \subref{fig:2D-ring} and spherical shell \subref{fig:3D-shell}. The magnetization distributions (green arrows) are obtained by means of micromagnetic simulations for permalloy samples \cite{Note3}. Nanorings have the same inner radius $R_{i}=15$~nm and outer radius $R_o=20$~nm, the thickness is $h=5$~nm and $h=1$~nm for the vortex-state (left) and onion-state (right) rings, respectively; vortex-state spherical shell (left) has inner radius $R=15$~nm and thickness $h=20$~nm, the onion-state shell (right) -- $R=24$~nm, $h=1$~nm. Color of the nanoparticle surfaces corresponds to the radial component of magnetization. Magnetizations of both poles are unidirectional.}
		\label{fig:topologicalCharges}
	\end{figure}

	The modern tendency is to extend flat structures into three dimensional (3D) space: the mutual cooperation between topology and curved geometry results in a rich physics as well as in a novel functionality, forming a new topic of a magnetism in curved geometries, for a review see Ref.~\cite{Streubel16a}. A thin spherical shell can be considered as one of the simplest 3D object, a bridgehead for studying the interplay of topological structures with a curvature of underlying surface. In order to elucidate the problem, we recall some topological issues for magnetization distribution in narrow magnetic rings as a 2D counterpart of this 3D object. The magnetization structure of nanorings is well--known \cite{Klaui03a, Kravchuk07, Guimaraes09, Sheka15} to form vortex and onion equilibrium states. The stability of non-trivial magnetization configuration can be explained by means of topological reasons. In the case of a ring, the topological properties of a planar magnetization distribution, $m_x+im_y=\exp(i\phi)$, on a closed loop $\gamma$ can be described by the $\pi_1$--topological charge, a vorticity (or a winding number), $q=1/(2\pi)
	\int_\gamma\!\mathrm{d}\phi \in \mathbb{Z}$. This results in $q = 1$ for the vortex state and $q = 0$ for the onion state, see Fig.~\ref{fig:2D-ring}. Therefore these magnetization states belong to the different homotopy classes. As a result the vortex state can not be continuously transformed into the onion state and vice versa if the magnetization remains in the plain of the ring. Separated by the energy barrier, the transformation from the onion to the vortex state occurs for the narrow rings as the first order phase transition~\cite{Sheka15}. 
	
	Now we consider the spherical shell: due to the additional space dimension, one can remove the topological difference between the onion and the vortex states. Moreover the onion state of the spherical shell can be considered as a limit case of the vortex state, see Fig.~\ref{fig:3D-shell}. According to the Poincare--Hopf theorem the magnetization of a spherical shell can not be everywhere tangential to the shell surface, even for the case of a strong easy-surface anisotropy. Thus the double-vortex state with two diametrically opposite vortex cores appears \cite{Milagre07,Kravchuk12a}. The topological properties of a 3D vector field $\vec{m}$ on a closed surface $S$ are determined by the $\pi_2$--topological charge $Q=1/(4\pi)\int_S\! \mathcal{J}\mathrm{d}S \in \mathbb{Z}$ (a skyrmion number) with $\mathcal{J}$ being the  the mapping Jacobian \cite{Dubrovin85p2}.  The skyrmion number depends on mutual polarities of the vortex cores \cite{Kravchuk16}: $Q=\pm1$ for the same polarities (both cores are magnetized inward or outward the sphere) and $Q=0$ for the opposite polarities. For the magnetically soft spherical shell the state with $Q=0$ is always energetically preferable \cite{Kravchuk12a}. Thus we limit ourself with considering only vortex state with opposite cores polarities, see~Fig.~\ref{fig:3D-shell}. The vortex and onion magnetization configurations belong to the same homotopy class and can be transformed one into other in a continuous way. 
	
	In addition to fundamental reasons, an interest to magnetic spherical shells is stimulated by experimental advanced in production of spherical hollow nanoparticles (spherical shells) as artificial materials with unusual characteristics and numerous applications \cite{Berkowitz04,Tartaj03, Ye07, Hu08, Cabot09, Simeonidis11, Sarkar12, Gong14, Sarkar15}. A variety of equilibrium magnetization configurations for spherical shells were identified in micromagnetic simulations \cite{Kong08a, Cabot09, Simeonidis11, Kravchuk12a} and interpreted based on experiments \cite{Ye07, Cabot09, Simeonidis11, Sarkar12, Gong14, Sarkar15, Ye10}. Different theoretical models predicted dissimilar equilibrium states \cite{Shute00, Goll04, Goll06, Milagre07, Kong08a, Kravchuk12a}. In particular, according to Goll \emph{et al.} \cite{Goll04,Goll06} homogeneous (monodomain), two-domain, four-domain, and vortex states can be energetically preferable for spherical nanoshells depending on geometrical and materials parameters. Micromagnetic simulations by \citet{Kong08a} testified three different states at the phase diagram: homogeneous, vortex, and onion states. 
	
	The purpose of the current study is to systematize possible equilibrium magnetization states and to construct the phase transition theory, which describes the transformation of magnetization states with varying material and geometrical parameters. We consider magnetically soft spherical shells of various radii and thicknesses. As we see below, the equilibrium magnetization state of a very thin shell is the onion one \cite{Kravchuk12a}: it results from the exchange interaction, which prefers the more homogeneous onion state with magnetization oriented in meridian directions. As opposed the magnetostatic interaction prefers the double-vortex state with the magnetization oriented in the parallels directions, since it minimizes the volume magnetostatic charges. That is why the ground state magnetization of the rigid sphere forms the vortex configuration \cite{Kim15a}. The competition between the exchange and volume magnetostatic contributions results in a second order phase transition between these two states: this is the subject of the current paper. 
	
	The paper is organized as follows. In Sec.~\ref{sec:model} we consider a model of the spherical shell and discuss possible equilibrium states. The theoretical description of the phase transition is presented in Sec.~\ref{sec:PhD}; we compare theoretical results with micromagnetic simulations. In section \ref{sec:discussion} we present final remarks and discuss possible perspectives. Some details concerning the energy calculation are presented in Appendix \ref{sec:energy}. The critical behavior is detailed in Appendix \ref{sec:critical}.

	
	\section{The Model of a thin magnetic shell: onion and double-vortex solutions}
	\label{sec:model}
	
	\begin{figure}
		\includegraphics [width=1.\columnwidth, angle=0]{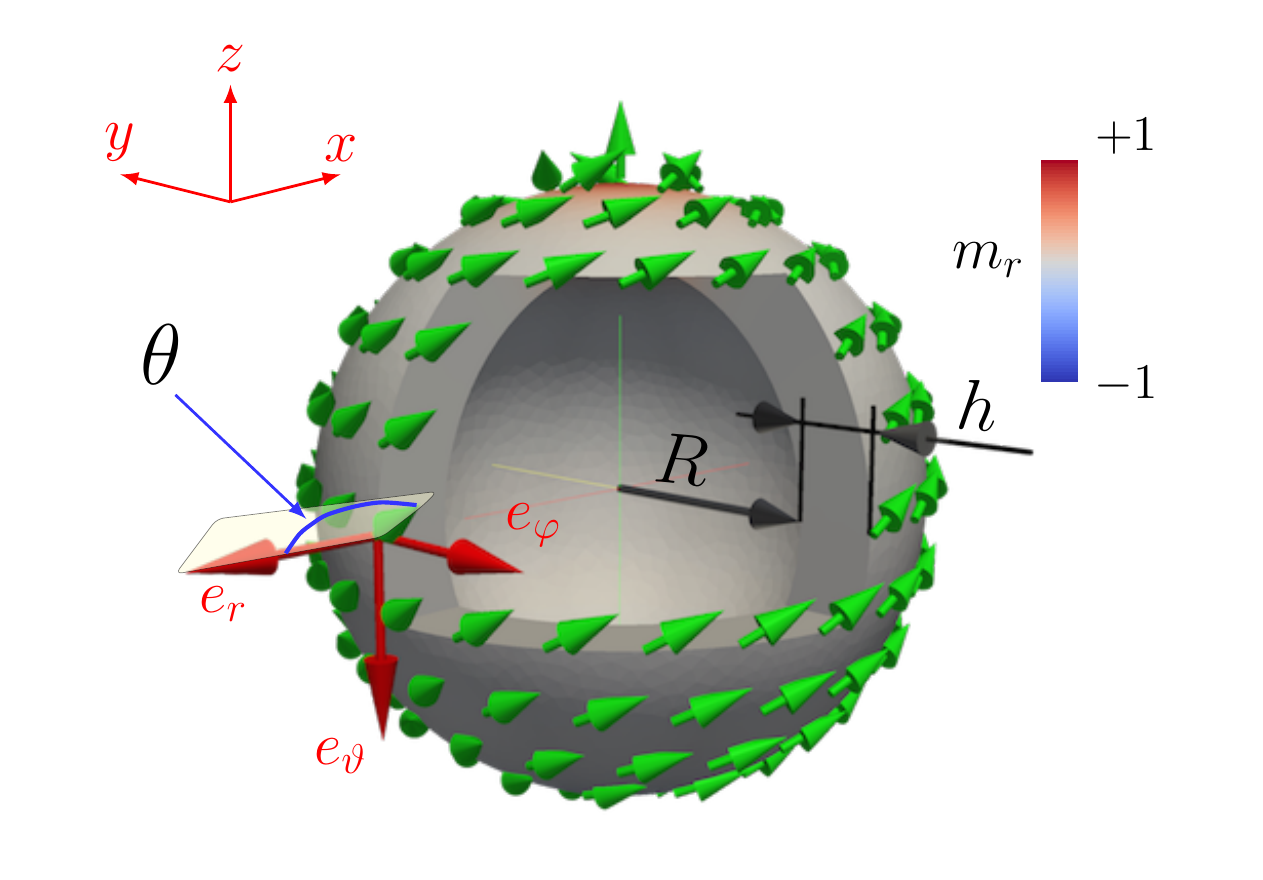}
		\caption{(Color online) 
			\textbf{Spherical shell: schematics and notation.} Green arrows correspond to the magnetization distribution.}
		\label{fig:schema}
	\end{figure}
	
	We consider a classical magnetically soft ferromagnet, using the continuous description for the unit magnetization vector $\vec{m}$. The minimal model takes into account two main interactions, which are described by the exchange energy $\mathscr{E}^{ex}$ and the magnetostatic one $\mathscr{E}^{ms}$. The total energy, normalized by $E_0 = 4 \pi M_s^2 V$ have the following form
	\begin{equation} \label{eq:energy}
	\begin{split}
	\mathscr{E} &= \mathscr{E}^{\text{ex}} +  \mathscr{E}^{\text{ms}},\\
	\mathscr{E}^{\text{ex}} &= -\frac{\ell^2}{2V} \int_V \mathrm{d}\vec{r} \left(\vec{m} \cdot \vec{\nabla}^2 \vec{m}\right),\\
	\mathscr{E}^{\text{ms}} &= \frac{1}{8\pi V}\!\!\int_V \!\!\mathrm{d}\vec{r}\left(\vec{m}(\vec r)\!\cdot\!\!\vec{\nabla}\right)\!\!\int_V\!\!\!\mathrm{d}\vec{r}'\left(\vec{m}(\vec r')\!\cdot\!\!\vec{\nabla}'\right)\!\! \frac{1}{|\vec{r}-\vec{r}'|}.
	\end{split}
	\end{equation}
	Here $M_s$ is the saturation magnetization, $V$ is the characteristic volume, $\ell = \sqrt{A/4\pi M_s^2}$ is the exchange length, and $A$ is the exchange constant.
	
	Let us specify the geometry as a thin spherical shell (see Fig. \ref{fig:schema}) with inner radius $R$ and thickness $h$. In order to keep the constrain $|\vec{m}|=1$ we utilize the common angular parameterization of magnetization in local spherical frame of reference
	\begin{equation} \label{eq:magnParam}
	\vec{m} = \cos\theta\,\vec{e}_r+\sin\theta\cos\phi\,\vec{e}_\vartheta+ \sin\theta\sin\phi\,\vec{e}_\varphi,
	\end{equation}	
	see Fig.~\ref{fig:schema} for the notations description. Here the angular magnetic
	variables $\theta = \theta(\vec{r})$ and $\phi = \phi(\vec{r})$ describe the magnetization distribution with respect to the spherical coordinates $(r,\vartheta, \varphi)$
	of the radius-vector $\vec{r}$.
	
	We limit our consideration by solutions of $\pi_2$--topological class $Q=0$, which are energetically preferable in comparison with higher $Q$ \cite{Kravchuk12a}. The corresponding magnetization structures include topologically trivial homogeneous magnetization distribution, the onion configuration and the double--vortex state (the last consist of two out-of-surface vortices 
	with opposite polarities \cite{Kravchuk12a}). The high symmetry of these configurations can be taken into account by considering: (i) azimuthally symmetric solutions only, i.e. the magnetization does not depend on $\varphi$, (ii) the odd symmetry under the spatial inversion, $m_r(\pi-\vartheta)=-m_r(\vartheta)$, and (iii) the uniform distribution along the radial direction, i.~e. the magnetization independence on $r$ \footnote{Strictly speaking, the assumption that the magnetization does not depend on $r$ is valid for thin shell limit, $h\ll R$.}. Following the symmetry of the magnetization distribution, we limit ourselves by the \emph{two-parameter Ansatz}
	\begin{subequations} \label{eq:Ansatz}
		\begin{align} \label{eq:Ansatz-theta}	
			\theta(\vec{r}) &=
			\begin{cases*}
				\frac{\pi}{2}-f(\vartheta,\lambda), & when $\vartheta \in \left[0,\frac{\pi}{2}\right)$,\\
				\frac{\pi}{2}& when $\vartheta =\frac{\pi}{2}$,\\
				\frac{\pi}{2}+f(\pi-\vartheta,\lambda), & when $\vartheta \in \left(\frac{\pi}{2},\pi\right]$,
			\end{cases*}
			\\
			\label{eq:Ansatz-Phi}
			\phi(\vec{r}) &= \pi - \varPhi.
		\end{align}
		According to Eqs.~\eqref{eq:Ansatz-theta}, \eqref{eq:Ansatz-Phi} the coordinate dependence of the magnetization is determined only by the polar angle $\vartheta$. There are two variational parameters: the \emph{declination angle} $\varPhi$ describes the slope of the magnetization with respect the meridian direction and the \emph{core parameter} $\lambda$, which controls the range of out--of--surface magnetization distribution situated on the sphere poles. The shape of the out--of--surface profile is determined by the shape--function $f(\vartheta,\lambda)$, which satisfies the following conditions:
		\begin{equation} \label{eq:Anzats-f-bs}
		f(0,\lambda) = \frac{\pi}{2}, \qquad f\left(\frac{\pi}{2},\lambda \right) = 0.
		\end{equation}	
		Without loss of generality we suppose that the magnetization is directed outward at the north pole, $\vec{m}(\vartheta=0)=\hat{\vec{z}}$ and inward at the south pole. To specify the shape--function we use the \emph{exponential} profile
		\begin{equation} \label{eq:Anzats-f}
		f(\vartheta,\lambda) = \frac{\pi}{2}\left(\frac{e^{-\frac{\vartheta}{\lambda}} - e^{\frac{\vartheta-\pi}{\lambda}}} {1-e^{-\frac{\pi}{\lambda}}}\right).
		\end{equation}
	\end{subequations}
	The exponential shape--function is a generalization of well--known Feldkeller Ansatz \cite{Feldtkeller65}, which is widely used for the vortices in planar disks \cite{Hubert98,Kravchuk07}. 
	
	\begin{table}
		\begin{tabular}{l c l l} \hline\hline
			Magnetization state  & $\mathscr{E}_1$	& Declination angle & Core size \\ \hline
			double--vortex  					 & $\mathscr{E}_1<0$	& $|\varPhi_0|\in\left(0,\pi/2 \right]$	    & $\lambda_0\in(0,\infty)$ \\
			onion								 & $\mathscr{E}_1>0$	& $\varPhi_0=0$	    & $\lambda_0\in(0,\infty)$ \\
			homogeneous							 & $\mathscr{E}_1>0$ 	& $\varPhi_0=0$	    & $\lambda_0=\infty$ \\
			\hline\hline
		\end{tabular}
		\caption{\textbf{Equilibrium magnetization states in a soft spherical shell:} possible configurations and corresponding variational parameters of the model \eqref{eq:Ansatz}. For convenience we include the homogeneous configuration as a limit case of the onion state.}
		\label{tab:states}
	\end{table}
	
	The main merit of the Ansatz \eqref{eq:Ansatz} is the possibility to describe different equilibrium magnetization states, see Table \ref{tab:states}: 
	
	(i) The double--vortex state is realized for finite $\varPhi$ and $\lambda$. The parameter $\lambda$ describes the single vortex core size, varying from $\lambda=0$ for the pure in-surface vortex to $\lambda=\infty$ for the homogeneous distribution. When $\vartheta>\lambda$, the magnetization has mostly in--surface components, directed along the meridians ($\varPhi=0$) for the very thin shell \cite{Kravchuk12a} and along the parallels ($\varPhi=\pm \pi/2$) for the very thick shell (rigid magnetic sphere) \cite{Kim15a}.
	
	(ii) The onion state has no azimuthal magnetization components, $\vec{m}\cdot \vec{e}_\varphi=0$; in the main part of the shell the magnetization is oriented along the meridian direction, $\vec{m}=-\vec{e}_\vartheta$, and only in the vicinity of the poles ($\vartheta<\lambda$) there appear the radial component \cite{Kravchuk12a}. 
	
	(iii) The homogeneous magnetization configuration $\vec{m}=\hat{\vec{z}}$ formally corresponds to the limit case of the onion state when $\lambda\to\infty$, see Eq.~\eqref{eq:Anzats-f} and Appendix \ref{sec:critical} for details. 
	
	Being homotopically equivalent, all three configurations can be continuously transformed to each other. Therefore one can consider onion and homogeneous configurations as limit cases of the double--vortex state: the onion state corresponds to the limit $\varPhi\to 0$; additionally to get the homogeneous configuration one has to consider the limit $\lambda\to\infty$.

	
	\section{Phase diagram}
	\label{sec:PhD}
	
	\begin{figure*}
		\begin{center}	
			\begin{subfigure}{\textwidth}
				\includegraphics [width=\columnwidth,angle=0]{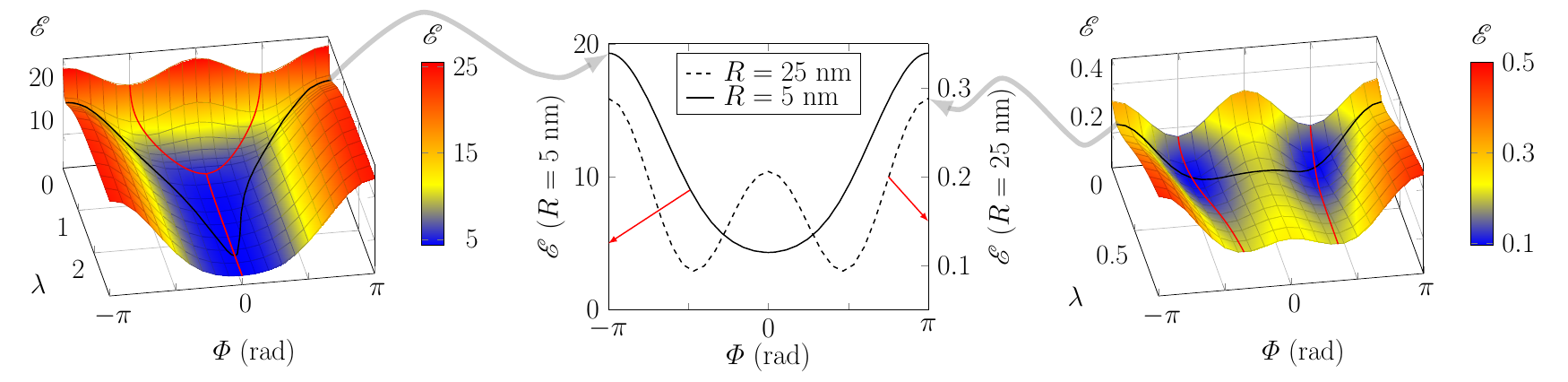}
			\end{subfigure}
			\begin{subfigure}{0.32\textwidth}
				\caption{$R=5$ nm, $h=10$ nm}
				\label{fig:energy_onion}
			\end{subfigure}
			\hfill
			\begin{subfigure}{0.32\textwidth}
				\caption{Energy \emph{vs} $\varPhi$ for equilibrium $\lambda$}
				\label{fig:energy2d}
			\end{subfigure}
			\hfill
			\begin{subfigure}{0.32\textwidth}
				\caption{$R=25$ nm, $h=10$ nm}
				\label{fig:energy_vortex}
			\end{subfigure}
			
			\caption{(Color online) \textbf{The energy landscape of a 3D shell:} the total energy \eqref{eq:TotalEnergyFinal} as a function of variational parameters $\varPhi$ and $\lambda$ for spherical shells with different geometrical parameters and the shape--function \eqref{eq:Anzats-f}.  Solid black and red lines correspond to the equilibrium values for variation parameters $\lambda$ and  $\varPhi$ respectively. Intersection of both lines means global energy minimum. The permalloy material parameters \cite{Note3} are used.}
			\label{fig:totalEnergy}
		\end{center}
	\end{figure*}
	
	Let us consider the energetics of different states. We apply the two-parameter Ansatz \eqref{eq:Ansatz} to the general energy expression \eqref{eq:energy}: The exchange contribution can be derived using the recent approach \cite{Gaididei14,Sheka15} for the arbitrary curved shell. One can compute the magnetostatic energy using the Legendre polynomials technique in the way similar to the magnetostatic energy calculation for monodomain state in hemispherical caps \cite{Streubel12,Sheka13b} and two-domain state in spherical shell \cite{Goll04, Goll06}. Finally the energy reads, see the Appendix \ref{sec:energy} for details:
	\begin{equation} \label{eq:TotalEnergyFinal}
	\begin{split}
	\mathscr{E}(\varepsilon, w; \lambda,\varPhi) &= \mathscr{E}^{\text{on}}(\varepsilon, w; \lambda) + \mathscr{E}_1(\varepsilon, w; \lambda) \sin^2\left(\frac{\varPhi}{2}\right)\\
	& + \mathscr{E}_2(\varepsilon; \lambda) \sin^4\left(\frac{\varPhi}{2}\right).
	\end{split}
	\end{equation}
	The energy depends on the geometrical parameter (the aspect ratio $\varepsilon = h/R$) and the reduced exchange length $w=\ell/R$. Besides, the energy is a function of variational parameters: the declination angle $\varPhi$ and the core parameter $\lambda$. Analysis shows that the energy term $\mathscr{E}_1$ results from the competition of exchange interaction and the  magnetostatic one: it takes the positive value when the exchange contribution dominates and the negative one when the magnetostatic plays a key role, see Eq.~\eqref{eq:E-on-1}. The last energy term, $\mathscr{E}_2$, does not depend on material parameter $w$ and always takes positive values, $\mathscr{E}_2>0$; the origin of this energy contribution is the volume magnetostatics only, see Eq.~\eqref{eq:E-2}.
	
	Equilibrium magnetization states can be found by minimization of the energy with respect to variational parameters. The equilibrium value of the declination angle~\footnote{To simplify the description we consider the case $\mathscr{E}_1>-2\mathscr{E}_2$. In opposite case one gets the onion solution with $\varPhi_0=\pm\pi$.}
	\begin{equation} \label{eq:Phi0}
	\varPhi_0 = 
	\begin{dcases*}
	0, & when $\mathscr{E}_1>0$,\\
	\pm 2\arcsin\sqrt{-\frac{\mathscr{E}_1}{2\mathscr{E}_2}}, & when $\mathscr{E}_1<0$.
	\end{dcases*}
	\end{equation}
	
	The magnetization distribution in the onion state is directed along the meridian, $\varPhi_0=0$. The energy of the onion state $\mathscr{E}^{\text{on}}(\varepsilon, w; \lambda)$ is determined mainly by the exchange energy and surface magnetostatic charges, see the Appendix \ref{sec:energy} for details. The onion state becomes energetically preferable under condition $\mathscr{E}_1>0$. The typical energy landscape $\mathscr{E}(\varPhi,\lambda)$ is an one--well potential with a well pronounced minimum at $\varPhi_0=0$ and some finite $\lambda_0>0$, see Figs.~\ref{fig:energy_onion}, \ref{fig:energy2d}.
	
	The double-vortex state with finite $\varPhi_0\neq0$ becomes energetically preferable when $\mathscr{E}_1<0$. The energy of the double-vortex state $\mathscr{E}^{\text{v}} = \mathscr{E}^{\text{on}} - \mathscr{E}_1^2/\left(4\mathscr{E}_2\right)$ is determined by both exchange and magnetostatic contributions. In this case we get a double--well energy landscape, see Figs.~\ref{fig:energy_vortex}, \ref{fig:energy2d}: both minima of equal depth are symmetrical in the direction of the declination angle $\varPhi$. That is why two different double-vortex states with opposite chiralities are energetically equivalent. 
	
\begin{figure}
		\includegraphics [width=1.\columnwidth, angle=0]{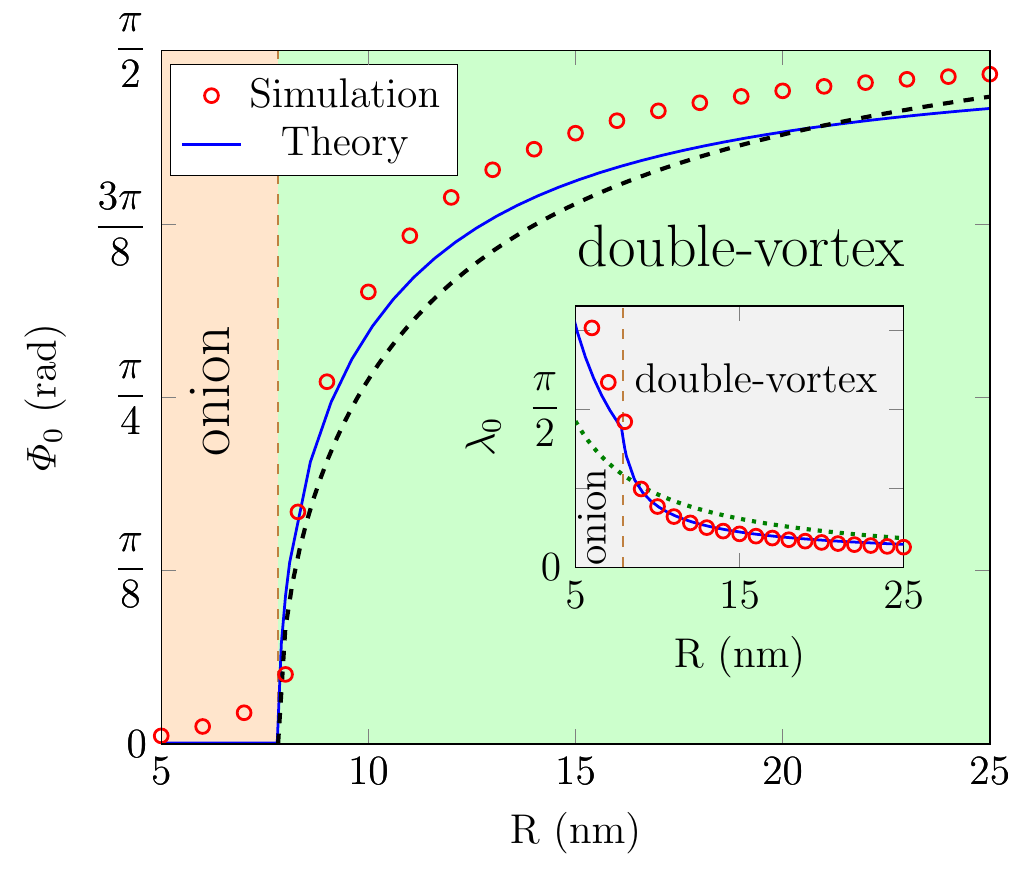}
		\caption{(Color online) \textbf{Equilibrium values of the variational parameters.} The declination angle $\varPhi_0$ as a function of inner radius for spherical shell with thickness $h=10$~nm. A solid line corresponds to the numerically calculated values for Ansatz \eqref{eq:Anzats-f}. Red circles correspond to the simulation data for the permalloy parameters. A black dashed line is obtained from the weakly nonlinear analysis (\ref{eq:nonlinear}). The inset shows the equilibrium core parameter $\lambda_0$ as a function of the inner radius: the dotted green line corresponds~to~$\lambda^{\text{disk}}$.}
		\label{fig:phiVsRadius}
	\end{figure}

	The equilibrium value of the core parameter $\lambda_0$ corresponds to the energy minimum, $\partial_\lambda \mathscr{E}(\varepsilon, w; \lambda,\varPhi_0)\bigr\rvert_{\lambda=\lambda_0}=0$, which can be found numerically only. Specifically, using the numerical minimization procedure for the energy \eqref{eq:TotalEnergyFinal}, we compute the equilibrium values $\varPhi_0$ and $\lambda_0$
	in a wide range of $R\in[0,30]$~nm and $h\in(0,30]$~nm. The onion state is found to be realized for spherical shells with either small radii or small enough thicknesses. The equilibrium declination angle $\varPhi_0$ rapidly increases with radius, reaching the limit value $\varPhi_0=\pi/2$ for large radii. The typical dependence $\varPhi_0(R)$ is shown in Fig.~\ref{fig:phiVsRadius} for the fixed thickness $h=10$~nm and the optimized core parameters $\lambda_0(R,h)$ using the Permalloy material parameters~\footnote{Permalloy is chosen as a material with the following parameters: the exchange constant $A = 21$~pJ/m, the saturation magnetization $M_s = 795$~kA/m. These parameters result in the exchange length $\ell\approx 5.14$~nm. Thermal effects and anisotropy are neglected.}. The onion state, $\varPhi_0=0$, is energetically preferable for the small enough radii, $R < R_c(h)$ with $R_c\approx 8$~nm for the chosen value $h=10$~nm. When $R>R_c$, the double-vortex state is realized, which is characterized by finite values of $\varPhi_0>0$, see Fig.~\ref{fig:phiVsRadius}.
	
	In order to verify the numerical results based on the model \eqref{eq:Ansatz} we perform a series of micromagnetic simulations in a wide range of radii $R\in[1,25]$~nm and shell thickness $h=10$~nm. For the simulations we used \textsf{magpar} code~\cite{MAGPAR, Scholz03a} with Permalloy material parameters \cite{Note3}. Equilibrium states are determined using numerical energy minimization starting from the double-vortex and homogeneous configurations. According to the simulations the angle $\varPhi$ practically does not depend on $\vartheta$, this is in agreement with the model assumption Eq.~\eqref{eq:Ansatz-Phi}. We consider the value of $\pi-\phi$ at the sphere equator as the equilibrium value $\varPhi_0$. The angles $\varPhi_0$ obtained in this way using simulations are shown in the Fig.~\ref{fig:phiVsRadius} by red circles. The good agreement with the model predictions should be noted. The core parameter $\lambda$ also demonstrates a good agreement with the model predictions, see the inset in the Fig.~\ref{fig:phiVsRadius}. Let us make a link with the planar disk, where the vortex core size $l^{\text{disk}} \approx \sqrt{2}\ell$ \cite{Feldtkeller65} under condition $h\ll\ell$ \cite{Kravchuk07}. This expression formally corresponds to the value $\lambda^{\text{disk}} \equiv l^{\text{disk}}/R=w\sqrt2$, which provides a good enough estimation for the vortex core size in a spherical shell, see inset in Fig.~\ref{fig:phiVsRadius}.
	
\begin{figure*}[ht]
		\begin{center}			
			\includegraphics [width=1.\textwidth, angle=0]{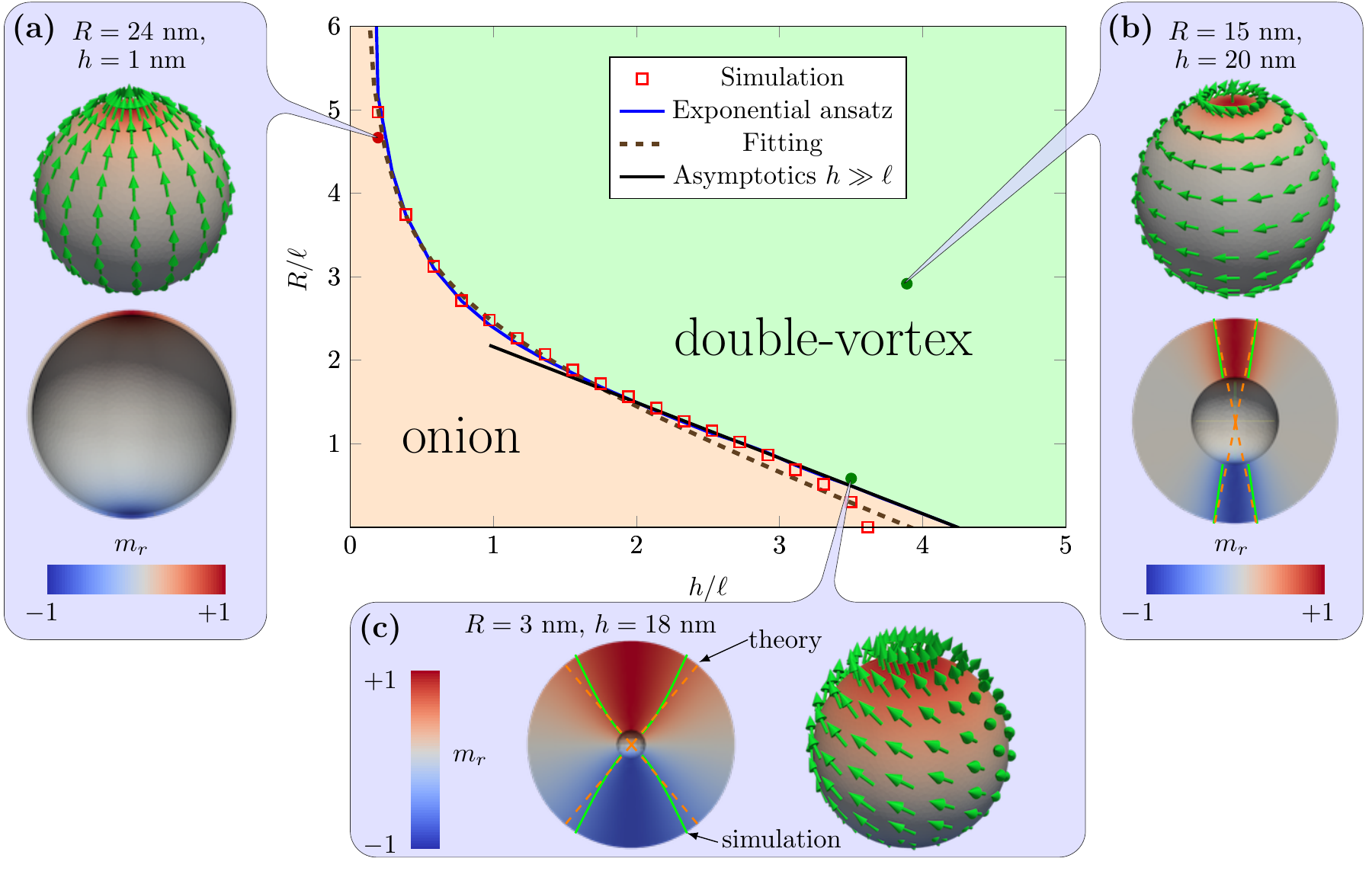}
			\caption{(Color online) \textbf{Phase diagram of equilibrium magnetization structures in the spherical shell.} Symbols correspond to the border between the onion and the double-vortex states obtained using simulation data. Solid blue line corresponds to theoretically calculated phase boundary as a numerical solution of \eqref{eq:critical} with account of the two-parameter  Ansatz \eqref{eq:Ansatz}. Solid black line corresponds to the asymptote \eqref{eq:RadiusVsThicknessForLargeEll}. Dashed brown line corresponds to the fitting \eqref{eq:fitting}. Outsets (a)--(c) show the equilibrium magnetization configurations for specified radii and thicknesses. Solid green and dashed orange curves are the isolines $m_r=0.7$ obtained from micromagnetic simulations and analytics, respectively.}
			\label{fig:phaseDiagram}
		\end{center}
	\end{figure*}

	Now we summarize results on equilibrium magnetization distributions. By comparing energies of
	different states, one can calculate the energetically preferable states for different geometrical parameters of the shell: its inner radius $R$ and the shell thickness $h$. 
	Simulations data are reproduced in Fig.~\ref{fig:phaseDiagram}. The general properties of the phase diagram are as follows. The equilibrium magnetization distribution of the very thin shell corresponds to the onion state in agreement with previous studies \cite{Kong08a,Kravchuk12a}. As opposed to this case the ground state of the large enough rigid sphere ($R=0$ and $h\gg\ell$) is the vortex state \cite{Goll04,Boardman05,Kim15a}. For the finite shell radii the transition from the onion state to the vortex one occurs as the thickness increases. In order to compute the  boundary between two phases we build numerically the dependence $\varPhi_0(R)$ in the way described above for the range of thicknesses $h\in[1;25]$~nm with step $\Delta h=1$~nm \footnote{In the vicinity of the transition point we choose $\Delta h=0.1$~nm.}. The critical value of the radius $R_c(h)$ is determined as the inflection point, $\partial_R^2\varPhi_0(R,h)\Bigr\rvert_{R=R_c}=0$. The obtained values $R_c$ are shown in Fig.~\ref{fig:phaseDiagram} by open squares.  
	
	Let us describe theoretically the critical behavior. The equilibrium declination angle $\varPhi_0$ takes zero values in the onion state, which corresponds to $\mathscr{E}_1>0$. It rapidly increases with radii following the critical square root behavior, see Fig.~\ref{fig:phiVsRadius}. Finally we get the finite value of $\varPhi_0$ in the double--vortex state, $\mathscr{E}_1<0$. Therefore we expect the second order phase transition at $\mathscr{E}_1=0$. We base our theoretical treatment of the phase transition on the energy approach. To derive the critical behavior we expand the energy \eqref{eq:TotalEnergyFinal} in series on $\varPhi$ at $\varPhi=0$:
	\begin{equation} \label{eq:Phase-Trans}
	\mathscr{E}= \mathscr{E}^{\text{on}} + \mathscr{E}_1 \frac{\varPhi^2}{4} + \left(\mathscr{E}_2 - \frac{\mathscr{E}_1}{3}\right) \frac{\varPhi^4}{16} +\mathcal{O}\left(\varPhi^6\right).
	\end{equation}
	It is important to emphasize that the $\varPhi^4$-term in Eq.~\eqref{eq:Phase-Trans} is always positive when $\mathscr{E}_1<0$, i.~e. in the double-vortex phase. This means the stability of the double-vortex solution. In the onion phase the energy term $\mathscr{E}_1$ becomes positive. At the transition point $\mathscr{E}_1=0$ one has $\mathscr{E}^{\text{on}}=\mathscr{E}^{\text{v}}$ and $\partial_\lambda\mathscr{E}^{\text{on}} = \partial_\lambda\mathscr{E}^{\text{v}}$. Thus the transition is continuous with respect to parameter $\lambda$.
	
	The boundary between two phases, i.~e. the critical curve, can be derived using the following conditions:
	\begin{equation} \label{eq:critical}
	\mathscr{E}_1\left(\varepsilon_c, w,\lambda_c\right)= 0,\qquad \partial_\lambda \mathscr{E}^{on}\left(\varepsilon_c, w, \lambda\right)\Bigr\rvert_{\lambda=\lambda_c} = 0.
	\end{equation}
	By excluding $\lambda{_c}$ from the set \eqref{eq:critical}, one obtains an equation for the critical curve $\varepsilon_c(w)$ which separates two phases in space of the geometrical parameters. This critical curve, recalculated in terms $R_c(h)$, is shown in Fig.~\ref{fig:phaseDiagram}. There is a good agreement with simulations data for the all range of parameters except $h>3\ell$, where the assumption about the magnetization uniformity along the thickness is violated, see Fig.~\ref{fig:phaseDiagram}(c).
	
	The critical curve can be calculated analytically in two limit cases. In the case of small radii one can use the uniform limit, which results in 
	\begin{equation} \label{eq:RadiusVsThicknessForLargeEll}
	\frac{R_c}{\ell} \approx  2\sqrt{2} - \frac{2}{3} \frac{h}{\ell},
	\end{equation}	
	see the Appendix \ref{sec:critical} for details; the asymptote \eqref{eq:RadiusVsThicknessForLargeEll} is shown by the solid line in Fig.~\ref{fig:phaseDiagram}. 
	
	Now we consider the opposite case of the large radii shells. The critical behavior is characterized as follows (see the Appendix \ref{sec:critical} for details):
	\begin{equation} \label{eq:R-infty}
	\frac{R_c}{\ell}  \approx 1.59 \left(\frac{\ell}{h}\right)^{2/3}.
	\end{equation}
	According to Eq.~\eqref{eq:R-infty} for each thickness $h$ there always exists a critical radius $R_c$, such that the onion state is energetically preferable one when $R<R_c$, and the vortex state has the lower energy when $R > R_c$.	
	
	Finally one can fit the boundary between the onion and the double--vortex states, obtained from the simulation data, in a wide range of parameters by the function
	\begin{equation} \label{eq:fitting}
	\frac{R_c}{\ell} \approx C_0 + C_1 \left(\frac{h}{\ell}\right)^{-\frac{2}{3}} - C_2\frac{h}{\ell},
	\end{equation}
	where $C_0 = 2.1$, $C_1 = 0.98$, $C_2 = 0.64$,  see the dashed brown line in the Fig.~\ref{fig:phaseDiagram}.
	
	Now being in possession of critical parameters, we are able to perform the weakly nonlinear analysis. At vicinity of the critical curve one can use the expansion \eqref{eq:Phase-Trans}. We consider the double-vortex phase near the critical parameters: $\varepsilon=\varepsilon_c$ and $w = w_c-\delta$ with $|\delta|\ll1$. Using the asymptotic behavior \eqref{eq:esp4w>>1} and \eqref{eq:lambda-eps-infty} for critical parameters we get
	\begin{equation} \label{eq:nonlinear}
	\begin{split}
	\varPhi_0 &\approx \psi_0\sqrt{1-\frac{w}{w_c}},\\
	\psi_0 = \frac{8 \varepsilon_c w_c^2 g_1(\lambda_c) }{\mathscr{E}_2(\varepsilon_c, w_c; \lambda_c)} &\approx
	\begin{cases*}
	\sqrt{15}, & when $w_c\ll1$,\\
	\sqrt{3}, & when $w_c\gg1$.
	\end{cases*}
	\end{split}	
	\end{equation}
	In terms of $R$ and $h$ this results in $\varPhi_0\propto \sqrt{1-R_c(h)/R}$,  which is well pronounced in the Fig.~\ref{fig:phiVsRadius}. In physics of nonlinear dynamic such behavior corresponds to the supercritical pitchfork bifurcation~\cite{Mayergoyz09}.
	
	
	\section{Discussion}
	\label{sec:discussion}
	
	Magnetic nanoparticles found a wide range of technological and biomedical applications \cite{Tartaj03,Gao09, Wu16}. Promising examples of effective combination of different materials and shapes include core/shell nanoparticles \cite{Wu16}, hemispherical Janus motors \cite{Baraban12}, and hollow magnetic spheres; they are widely used for drug delivery by magnetic field manipulation~\cite{Wang03, Cao08, Sarkar15}. Magnetic properties of nanoparticles and field response significantly depend on geometry. Therefore the systematic analysis of possible equilibrium magnetization states is of essential importance. In the current study performed such analysis and constructed the theory of phase transitions in the spherical shells of a soft nanomagnet. We found existence of two different magnetization states depending on geometrical parameters (the inner shell radius $R$ and thickness $h$), see the diagram of ground states in Fig.~\ref{fig:phaseDiagram}: (i) The onion state with the magnetization oriented mostly along meridians is typical for small enough thicknesses. (ii) The double--vortex state with the magnetization oriented mostly along parallels is preferable one in the opposite case. A second order phase transition in the space of geometrical parameters takes place between these two phases. All analytical results are verified by means of micromagnetic simulations. 
	
\begin{figure}
		\includegraphics [width=1.\columnwidth, angle=0]{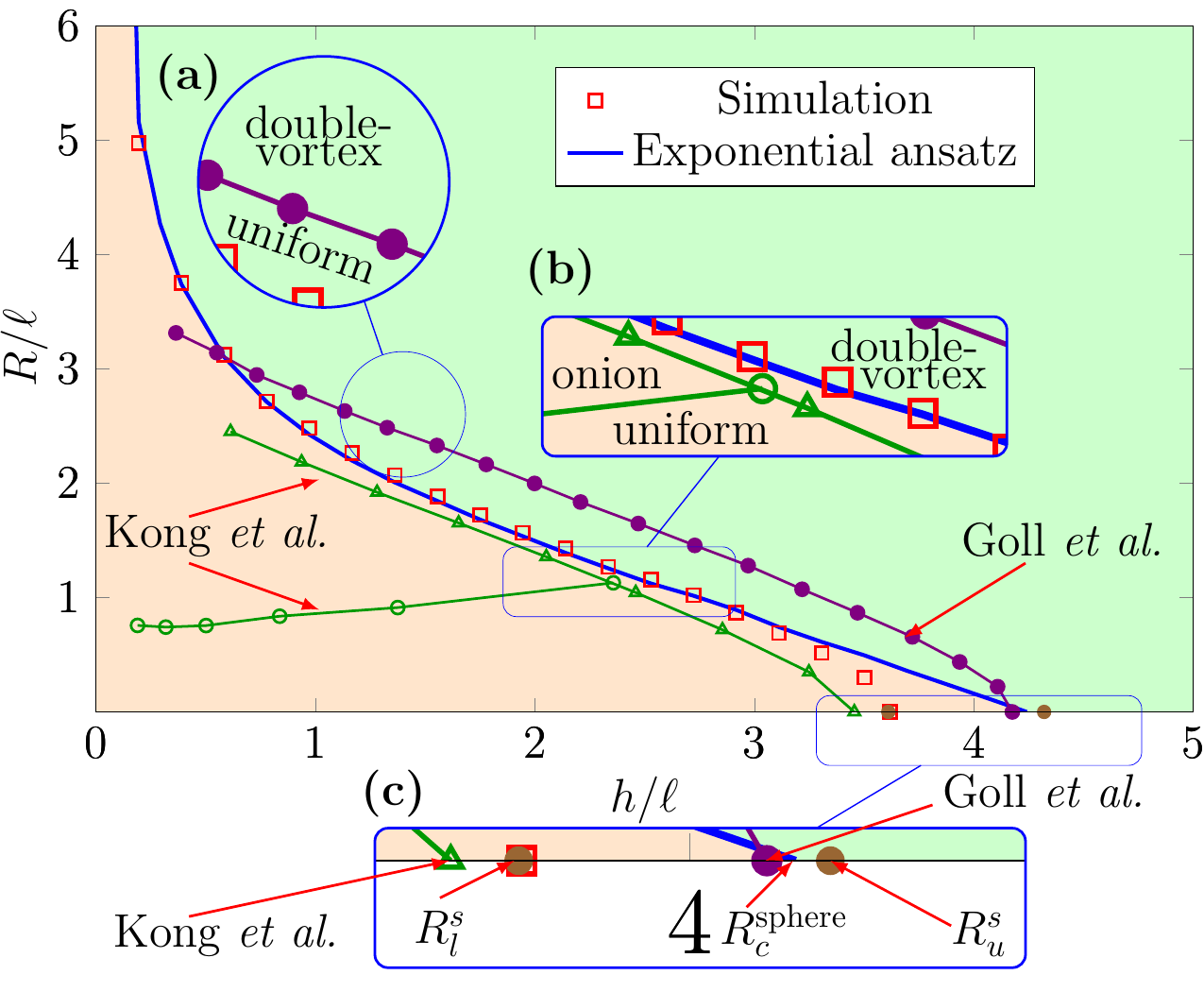}
		\caption{(Color online) \textbf{Phase diagram for different models.} Comparison of our results (blue line with open squares) with other models, see text for details: (a) violet line with filled circles corresponds to boundary between double-vortex and uniform states in~\citet{Goll04}; (b) green line with open circles and triangles corresponds to boundary between double-vortex, onion and uniform states in \citet{Kong08}. (c) The brown points correspond to the lower $R_l^s$ and $R_u^s$ bounds for the critical radius according to Brown's fundamental theorem~\cite{Brown68,Brown69,Aharoni96,Aharoni01}.
		}
		\label{fig:phaseDiagramComp}
	\end{figure}

	Let us compare results of this paper with previous studies. In particular, equilibrium magnetization states in spherical caps were investigated experimentally \cite{Ye07, Cabot09, Simeonidis11, Sarkar12, Gong14}, numerically \cite{Kong08a, Cabot09, Simeonidis11}, and theoretically \cite{Goll04,Goll06, Milagre07, Kong08a, Kravchuk12a}. Goll \emph{et al.} \cite{Goll04,Goll06} classified possible equilibrium magnetization distributions as  double-vortex state and strictly uniform one. However, it is well known \cite{Aharoni96} that the strictly homogeneous distribution is not allowed due to stray field effects, except the case of the rigid sphere. Underestimation of stray field energy effectively shifts the boundary between the monodomain state and the double-vortex one to the region of higher radii, see violet filled circles in Fig.~\ref{fig:phaseDiagramComp}(a); according to our analysis this curve lies in the double-vortex state. More essential, the crudeness of the model (not enough sensitive Ansatz) can drastically change the type of the phase transition: the first order transition according to Refs.~\cite{Goll04,Goll06} instead of the second order phase transition according to the current study. One has to mention the coincidence of results by \citet{Goll04} with our theoretical curve (blue line) at the limit case $R=0$, which corresponds to the rigid sphere with $R_c^{\text{sphere}}$, see Fig.~\ref{fig:phaseDiagramComp}(c). 
	
	More precise technique was used in Ref.~\cite{Kong08a}, where a planar ring symmetry consideration \cite{Landeros06} was  extended to a 3D spherical shell. More specifically, \citet{Kong08a} distinguished strictly uniform, onion and vortex states which results in a three-phases diagram, see green line in Fig.~\ref{fig:phaseDiagramComp}(b). The curve for the boundary with a double--vortex state is in a good agreement with out simulations, see red open squares and green open triangles in Fig.~\ref{fig:phaseDiagramComp}. Apart this curve, one more boundary was identified in Ref.~\cite{Kong08a}: the green line with open circles separates the onion and uniform states, see Fig.~\ref{fig:phaseDiagramComp}(b). This curve correspond  to sample size comparable with the exchange length, for which the exchange energy definitely dominates over all other contributions. That is why the magnetization distribution of small radius shell in the onion state was difficult to distinguish from the pure uniform~one.
	
	Our analysis verifies the continuous transition between double-vortex and onion state. To distinguish vortex and single-domain (onion) state we used the Ansatz~\eqref{eq:Ansatz} with a variable inclination angle $\varPhi$. This approach allows to get a boundary between different equilibrium magnetization states in good agreement with simulation data, see Fig.~\ref{fig:phaseDiagram}.

	It is instructive to discuss the limit case of a rigid sphere. In this limit Brown's fundamental theorem \cite{Brown68,Brown69,Aharoni96} provides lower and upper bounds for the critical radius: $R_{\text{l}}^{\text{s}} \approx 3.61\ell$ and the upper bound (corrected by \citet{Aharoni01}) $R_{\text{u}}^{\text{s}}\approx 4.32\ell$. When the sphere radius is smaller than the critical value, $R<R_c^{\text{s}} \in \left(R_{\text{l}}^{\text{s}}; R_{\text{u}}^{\text{s}}\right)$, the magnetization configuration forms a \emph{strictly homogeneous} monodomain state. The sphere can be considered as a limit case of spherical shell. Therefore one can consider the homogeneous state (valid for the sphere) as a limit case of the onion state (valid for the spherical shell). Let us apply the asymptotic result \eqref{eq:RadiusVsThicknessForLargeEll} to the limit case of the rigid sphere, $R=0$. Then the thickness of the shell $h$ determines the radius of the sphere with the critical value $R_c^{\text{sphere}}=3\sqrt{2}\ell\approx 4.24\ell$, which is closed to the critical value, obtained by micromagnetic simulations, $R_c^{\text{sphere}}\approx 3.62\ell$. This result is also in agreement with above mentioned Brown's bounds, see outset in Fig.~\ref{fig:phaseDiagramComp}.
	
	
	\section*{Acknowledgments}
	\label{sec:Acknowledgments}
	
	All simulations results presented in the work were obtained using the computing clusters of Taras Shevchenko National University of Kyiv \cite{unicc} and Bayreuth University \cite{btrzx}. V.P.K. acknowledges the Alexander von Humboldt Foundation for the support and IFW Dresden for kind hospitality.
	
	
	\appendix
	
	\section{Energy calculation details}
	\label{sec:energy}
	
	The purpose of this appendix is to derive the energy \eqref{eq:TotalEnergyFinal}. We use the energy functional \eqref{eq:energy} and apply the two-parameter Ansatz \eqref{eq:Ansatz}.
	
	Let us start with the exchange energy calculation. Recently the full 3D theory for thin magnetic shells of arbitrary shape was put forth in Ref.~\cite{Gaididei14,Sheka15}, see Ref.~\cite{Streubel16a} for a review. The normalized exchange energy $\mathscr{E}^{\text{ex}}$, see Eq.~\eqref{eq:energy}:		
	\begin{equation} \label{eq:Exchange-energy}
	\mathscr{E}^{\text{ex}} = \frac{\ell^2}{2V}\int_{\mathclap{R}}^{\mathclap{R+h}}\! \mathrm{d}r\,r^2\int_0^{2\pi}\!\!\mathrm{d}\varphi\int_{0}^{\pi}\!\! \mathrm{d}\vartheta\sin\vartheta\, W^{\text{ex}}[\theta,\phi],
	\end{equation}
	where the exchange energy density reads \cite{Gaididei14} 
	\begin{equation*} 
		W^{\text{ex}}[\theta,\phi] = \left[\vec{\nabla} \theta - \vec{\varGamma}\right]^2 + \left[\sin\theta \left(\vec{\nabla} \phi - \vec{\varOmega}\right) \! -  \! \cos\theta \partial_\phi \vec{\varGamma}\right]^2\!\!.
	\end{equation*}
	Considering the spherical geometry we normalize the energy using the sphere volume $V = (4/3)\pi R^3$; the tangential vector $\vec{\varGamma}(\phi) =-r^{-1}(\cos\phi\,\vec{e}_\vartheta+\sin\phi\,\vec{e}_\varphi)$ and the spin connection $\vec{\varOmega} = -r^{-1}\cot \vartheta \vec{e}_{\varphi}$. 
	
	Now we substitute the Ansatz \eqref{eq:Ansatz} and perform the integration over $r$ and $\varphi$, which results in
	\begin{equation} \label{eq:Exchange-energy-1}
	\mathscr{E}^{\text{ex}}(\varepsilon,w; \lambda, \varPhi) =
	w^2\varepsilon \left[ g_0(\lambda) - g_1(\lambda)  \cos\varPhi \right].
	\end{equation}	
	Here $g_0$ and $g_1$ are determined by the profile of the shape--function $f(\vartheta,\lambda)$ in the following way
	\begin{equation*} \label{eq:gk}
		\begin{split}
			g_0(\lambda) = \frac{3}{2} \int_{0}^{\pi/2} \mathrm{d} \vartheta \sin\vartheta \Bigl[ 
			&(\partial_\vartheta f(\vartheta,\lambda) )^2+ 1 \\
			&\left.+\sin^2 f(\vartheta,\lambda) +\cot^2\vartheta \cos^2 f(\vartheta,\lambda) \right],\\
			g_1(\lambda) = 3 \int_{0}^{\pi/2} \mathrm{d} \vartheta \Bigl[& \cos\vartheta \sin f(\vartheta,\lambda) \cos f(\vartheta,\lambda)  \\
			&- \sin \vartheta ~\partial_\vartheta f(\vartheta,\lambda) \Bigr].
		\end{split}
	\end{equation*}
	
	In order to calculate the normalized magnetostatic energy $\mathscr{E}^{\text{ms}}$, see Eq.~\eqref{eq:energy}, we utilize the expansion
	\begin{equation*}
		\begin{split}
			\frac{1}{\left| \vec{r} - \vec{r}' \right|} &=
			\frac{1}{r_>} \sum_{l=0}^{\infty} \sum\limits_{m=-l}^{l} \left( \frac{r_<}{r_>}\right)^l \frac{(l-m)!}{(l+m)!}\\
			&\times P_l^m(\cos\vartheta)  P_l^m(\cos\vartheta ') e^{im(\varphi - \varphi')},
		\end{split}
	\end{equation*}
	where $r_<=\min(r,r')$, $r_>=\max(r,r')$, and $P_l^m(\cos\vartheta)$ is the associated Legendre polynomial \cite{NIST10}. Substituting now the Ansatz \eqref{eq:Ansatz} into $\mathscr{E}^{\text{ms}}$ and performing the integration over space coordinates one obtains the expression 
	\begin{equation} \label{eq:finalMagnetostaticEnergy}
	\begin{split}
	\mathscr{E}^{\text{ms}}(\varepsilon; \lambda,\varPhi) &= \mathscr{E}_0^{\text{ms}}(\varepsilon; \lambda) - \mathscr{E}_1^{\text{ms}}(\varepsilon; \lambda) \cos\varPhi\\
	&+  \mathscr{E}_2^{\text{ms}}(\varepsilon; \lambda) \cos^2\varPhi.
	\end{split}
	\end{equation}
	Here $\mathscr{E}_0^{\text{ms}}$ and $\mathscr{E}_2^{\text{ms}}$ originate from surface and volume effective magnetostatic charges, respectively, and $\mathscr{E}_1^{\text{ms}}$ represents the interaction of these charges.

	Due to the high symmetry of the two-parameter Ansatz the energy contributions has a relatively simple form
	\begin{equation}\label{eq:Ems_n}
	\begin{split}
	\mathscr{E}_0^{\text{ms}}(\varepsilon; \lambda) &= \frac{3}{2}\sum\limits_{l=1}^\infty \mathcal{F}_l(\varepsilon)\mathcal{A}_l^2(\lambda),\\
	\mathscr{E}_1^{\text{ms}}(\varepsilon; \lambda) &= -\frac{3}{2}\sum\limits_{l=1}^\infty \mathcal{G}_l(\varepsilon)\mathcal{A}_l(\lambda)B_l(\lambda),\\ 
	\mathscr{E}_2^{\text{ms}}(\varepsilon; \lambda) &= \frac{3}{2}\sum\limits_{l=1}^\infty \mathcal{G}_l(\varepsilon) \mathcal{B}_l^2(\lambda).
	\end{split}
	\end{equation}
	The shape-function determines the values of $\mathcal{A}_l(\lambda)$ and $\mathcal{B}_l(\lambda)$: only odd harmonics survive for the symmetry reason:
	\begin{align*}
		\mathcal{A}_l(\lambda) &= \frac{1-(-1)^l}{2}\int\limits_{0}^{\pi/2}\sin\vartheta\sin f(\vartheta,\lambda)P_l(\cos\vartheta)\mathrm{d}\vartheta,\\
		\mathcal{B}_l(\lambda) &= \frac{1-(-1)^l}{2}(l+1)\int\limits_{0}^{\pi/2}\cos f(\vartheta,\lambda)\\
		&\times\left[P_{l+1}(\cos\vartheta)-\cos\vartheta P_l(\cos\vartheta)\right] \mathrm{d}\vartheta.
	\end{align*}
	Parameters $\mathcal{F}_l$ and $\mathcal{G}_l$ depend on the aspect ratio $\varepsilon$:
	\begin{align*}
		\mathcal{F}_l(\varepsilon) &= \frac{1}{l+2}\left[(3l+2)\alpha(\varepsilon)+2l(l+1)\beta_l(\varepsilon)\right],\\
		\mathcal{G}_l(\varepsilon) &= \frac{2}{l+2}\left[\alpha(\varepsilon)-\beta_l(\varepsilon)\right],
	\end{align*}
	where we use the following notations
	\begin{equation*}
		\begin{split}
			\alpha(\varepsilon) &= \frac13\left[(1+\varepsilon)^3-1\right],\\
			\beta_l(\varepsilon) &=
			\begin{dcases*}
				\ln(1+\varepsilon), & when $l=1$,\\
				\frac{1}{l-1}\left[1-\frac{1}{(1+\varepsilon)^{l-1}}\right], & when $l>1$.
			\end{dcases*}
		\end{split}
	\end{equation*}
	Note that $\mathcal{G}_l\approx\varepsilon^2$ and $\mathcal{F}_l\approx \varepsilon\left[2(l+1)^2+l\right]/(l+2)$ for the case $\varepsilon\ll1$. It is important to stress a very rapid convergence of series \eqref{eq:Ems_n} on $l$: in our analysis we limit ourselves by terms with $l=1$ only, which provides the accuracy of the magnetostatic energy calculation of about 6\%.
	
	Now we sum up the information about different energy contributions. According to Eq.~\eqref{eq:energy}, the total energy is a combination of the exchange energy \eqref{eq:Exchange-energy-1} and the magnetostatic one \eqref{eq:Ems_n}:
	\begin{equation} \label{eq:TotalEnergyFinal-1}
	\tag{\ref{eq:TotalEnergyFinal}$'$}
	\mathscr{E} = \mathscr{E}^{\text{on}} + \mathscr{E}_1 \sin^2(\varPhi/2) + \mathscr{E}_2 \sin^4(\varPhi/2).
	\end{equation}
	Here $\mathscr{E}^{\text{on}}$ and $\mathscr{E}_1$ depend on geometrical parameter $\varepsilon$, the material parameter $w$, and the variational core parameter~$\lambda$:
	\begin{subequations} \label{eq:E-on-1-2}
		\begin{equation} \label{eq:E-on-1}
		\begin{split}
		\mathscr{E}^{\text{on}}(\varepsilon, w; \lambda) &= w^2 \varepsilon \left[g_0(\lambda)-g_1(\lambda)\right] + \mathscr{E}_0^{\text{ms}}(\varepsilon; \lambda)\\
		&- \mathscr{E}_1^{\text{ms}}(\varepsilon; \lambda) + \mathscr{E}_2^{\text{ms}}(\varepsilon; \lambda),\\
		\mathscr{E}_1(\varepsilon, w; \lambda) &= 2w^2 \varepsilon g_1(\lambda) + 2\mathscr{E}_1^{\text{ms}}(\varepsilon; \lambda)\\
		& -4 \mathscr{E}_2^{\text{ms}}(\varepsilon; \lambda).
		\end{split}
		\end{equation}	
		The last term $\mathscr{E}_2$ is not affected by the reduced exchange length $w$, this energy is caused by the magnetostatic contribution of the volume charges only:
		\begin{equation} \label{eq:E-2}
		\mathscr{E}_2(\varepsilon; \lambda) = 4\mathscr{E}_2^{\text{ms}}(\varepsilon; \lambda).
		\end{equation}
	\end{subequations}
	The energy term $\mathscr{E}_1$ appears as a competition of exchange interaction and the  magnetostatic one: in the case $w\gg1$ (small radii) the exchange contribution dominates, hence $\mathscr{E}_1>0$ and the onion state is realized. In the case $\varepsilon\gg1$ (thick shells) the contribution of the volume magnetostatic charges overcomes the exchange term, $\mathscr{E}_1<0$ and the double-vortex state becomes preferable.

	
	\section{The critical curves}
	\label{sec:critical}
	
	In order to compute the critical curves $\varepsilon_c(w)$ we use the set of Eqs.~\eqref{eq:critical}. The analysis can be done in two limit cases.
	
	In the limit case $w\gg1$ one gets a homogeneous magnetization distribution $\vec{m}=\hat{\vec{z}}$, which can be described by the shape--function
	\begin{equation} \label{eq:uniform}
	f(\vartheta,\lambda)=\frac{\pi}{2} - \vartheta.
	\end{equation}
	Formally, this corresponds to Eq.~\eqref{eq:Anzats-f} under the limit $\lambda\to\infty$. Proceeding in Eqs.~\eqref{eq:critical} to this limit, we get
	\begin{equation} \label{eq:esp4w>>1}
	\varepsilon_c  = 3\sqrt{2}w-\frac{3}{2} + \mathcal{O}\left(\frac{1}{w}\right).
	\end{equation}
	In terms of $R$ and $h$ this asymptote takes the form Eq.~\eqref{eq:RadiusVsThicknessForLargeEll}.
	
	In the opposite case $w\ll1$ the vortex core size $\lambda_c\ll1$, hence the shape--function \eqref{eq:Anzats-f} becomes singular. That is why to consider the critical behavior for the large radii we modify the shape--function $f(\vartheta,\lambda)$ using the \emph{linear} profile
	\begin{equation} \label{eq:linear-Ansatz}
	f(\vartheta,\lambda) = 
	\begin{cases}
	\frac{\pi}{2}\left(1-\frac{\vartheta}{\lambda}\right), & \text{when $\vartheta \in [0, \lambda)$},\\
	0, & \text{when $\vartheta \in \left[\lambda,\frac{\pi}{2}\right)$}.
	\end{cases}
	\end{equation}
	In the main approach on the small $w\ll1$ we get the following asymptote behavior		
	\begin{equation} \label{eq:lambda-eps-infty}
	\begin{split}
	\lambda_c &= 1.12 \sqrt{w} + \mathcal{O}\left(w^{3/2}\right), \\
	\varepsilon_c &= 2 w^{5/2} + \mathcal{O}\left(w^{7/2}\right).
	\end{split}
	\end{equation}
	Finally, this results in Eq.~\eqref{eq:R-infty}.

	%
	%
	%
	%

	
\end{document}